\documentclass[%
 aip,
 amsmath,amssymb,
 reprint,%
nofootinbib]{revtex4-1}

\usepackage{graphicx}
\usepackage{dcolumn}
\usepackage{bm}

\usepackage[utf8]{inputenc}
\usepackage[T1]{fontenc}
\usepackage{mathptmx}
\usepackage{etoolbox}
\usepackage{amssymb} 
\usepackage{caption} 
\usepackage{enumitem} 

\makeatletter
\def\@email#1#2{%
 \endgroup
 \patchcmd{\titleblock@produce}
  {\frontmatter@RRAPformat}
  {\frontmatter@RRAPformat{\produce@RRAP{*#1\href{mailto:#2}{#2}}}\frontmatter@RRAPformat}
  {}{}
}%
\makeatother 

\usepackage[symbol]{footmisc}
\renewcommand{\thefootnote}{\fnsymbol{footnote}}

\usepackage{url}

\usepackage{tocloft}
\usepackage{xcolor}
\usepackage{soul}
\usepackage{comment}

\begin{document}

\preprint{AIP/123-QED} 

\title[Analyzing Machine Learning Performance in a Hybrid Quantum Computing and HPC Environment]{Analyzing Machine Learning Performance in a Hybrid Quantum Computing and HPC Environment}

\thanks{This manuscript has been authored by UT-Battelle, LLC under Contract No. DE-AC05-00OR22725 with the U.S. Department of Energy.  The publisher, by accepting the article for publication, acknowledges that the U.S. Government retains a non-exclusive, paid up, irrevocable, world-wide license to publish or reproduce the published form of the manuscript, or allow others to do so, for U.S. Government purposes. The DOE will provide public access to these results in accordance with the DOE Public Access Plan (http://energy.gov/downloads/doe-public-access-plan).}


\author{Samuel T. Bieberich}
\affiliation{Texas A\&M University, Wisenbaker Engineering Building 3128, 188 Bizzell St, College Station, TX 77843}

\author{Michael A. Sandoval}%
\affiliation{National Center for Computational Sciences, Oak Ridge National Laboratory, P.O. Box 2008, Oak Ridge, TN 37831-6164, USA}

\date{\today}

\begin{abstract}
We explored the possible benefits of integrating quantum simulators in a “hybrid” quantum machine learning (QML) workflow that uses both classical and quantum computations in a high-performance computing (HPC) environment.
Here, we used two Oak Ridge Leadership Computing Facility HPC systems, Andes (a commodity-type Linux cluster) and Frontier (an HPE Cray EX supercomputer), along with quantum computing simulators from PennyLane and IBMQ to evaluate a hybrid QML program -- using a "ground up" approach.
Using 1 GPU on Frontier, we found $\sim$56\% and $\sim$77\% speedups when compared to using Frontier's CPU and a local, non-HPC system, respectively.
Analyzing performance on a larger dataset using multiple threads, the Frontier GPUs performed $\sim$92\% and $\sim$48\% faster than the Andes and Frontier CPUs, respectively.
More impressively, this is a $\sim$226\% speedup over a local, non-HPC system's runtime using the same simulator and number of threads.
We hope that this proof of concept will motivate more intensive hybrid QC/HPC scaling studies in the future.
\end{abstract}

\maketitle


\section{Introduction}
Ever since Alan Turing's landmark paper "Computer Machinery and Intelligence" in 1950\cite{turing}, the field of computing has experienced unprecedented growth.
Seventy-three years later, High Performance Computing (HPC) machines, often called "supercomputers," operate on the Exascale (1 quintillion operations per second)\cite{exascale}, and Quantum Computing has morphed from a science fiction pipe dream to near-term reality\cite{supremacy}. 
With this growth in computing power, it is only natural to look back at the initial propositions of scientists like Turing and question whether computers can be programmed to "think" in the same way as the human mind.
While simulating the methods behind the most complicated biological construct on Earth often has been viewed as futility, the latest developments in Machine Learning have asserted that this is not the case with modern computing equipment.  

Machine Learning is a sub-field of Artificial Intelligence, or AI, a branch of computer science that studies the ability of computers to emulate the intellectual processes of human beings, such as (as per Britannica) "the ability to reason, discover meaning, generalize, or learn from past experience."\cite{britannica}
Common applications include in Google Search algorithms, Autocorrect, and traffic predictions. 

Machine Learning, and in particular the method we utilized most in this project, Convolutional Neural Networks (CNNs), allow computers to utilize data to recognize patterns that would otherwise require an immense amount of hard coding.
For example, one of the most famous uses of CNNs is in image recognition.
By creating what is called a "neural network" composed of various layers, each with different weights, linear algebra operations through the network allow the machine to recognize the pixel patterns appearing in different files and sort them into groups. 
After this is completed, accuracy is determined by reading the labels of all of the photos through a process called "supervised learning"\cite{pytorch-docs}.
The process is repeated by stepping backward and then forwards again through the network, manipulating weights until final accuracy across the widest range of data is optimized.  

While modern HPC infrastructure is currently the primary method for which ML programs are run, a promising option for the next era of the industry is Quantum Machine Learning (QML).
As datasets grow in the coming decades, the utilization of Quantum Computers (QC), often considered critical in the next generation of high-performance computing, is a natural progression of computing power that Machine Learning (ML) will require.
Quantum Computers specifically are well equipped to handle processes like those in CNNs, optimized for computing linear algebra matrix calculations and weighted cost functions. 
They also are much more scalable than current HPC infrastructure, slowed by the steady decline of Moore's Law\cite{moore-end}.
Given this over-idealized outlook, the question remains, why would HPC need to be involved at all? 

Quantum Computers can't solve all computable problems, and while classical computers share this fact, they have a much broader repertoire, easily explained by their 6 added decades of relevance and R\&D.
The lack of Turing Complete Quantum Computers prevents them from accurately simulating classical architecture, meaning that, at least for the foreseeable future, HPC and QC will be best utilized in tandem, rather than completely replacing one or the other\cite{meetiqm}. 

This project aims to confirm the modern use cases of hybrid quantum and classical machine learning algorithms and analyze the practicality of running said programs on real HPC hardware, serially and in parallel.
We ran tests from several programs on the computers at Oak Ridge Leadership Computing Facility (OLCF) at Oak Ridge National Lab, setting benchmarks and determining the current computational feasibility of hybrid HPC and QC programs in the field of ML.

\section{Baseline ML Tests on HPC resources}

To establish a suitable standard for evaluating hybrid algorithms, our initial step involved conducting multiple tests based on PyTorch's documentation, specifically the "Transfer Learning for Computer Vision Tutorial"\cite{pytorch-docs}.
PyTorch\cite{pytorch-citation} is a Python library that specializes in ML, offering various datasets and tools to allow for training using CNNs, among other methods, either locally on your personal computer or remotely on servers or HPC clusters. 
As stated before, CNNs are often used for image recognition algorithms, but oftentimes training a dataset is difficult when it is sufficiently small.
Transfer Learning takes this into account by taking the training from the top layers of a large CNN, and then re-appropriating it for a related task\cite{TL}. 
The final layers are then completed again, allowing for a relatively accurate and fast result (given the correct dataset).
For the first example explored until Section \ref{sec:exp-data}, the dataset is composed of approximately 240 images of ants and bees. 
Compared to ImageNet\cite{imagenet}, which has 14 million images, this would normally be not enough data for a computer to train off of.
However, since this is just a small subset of ImageNet, the PyTorch code uses the top layers from the main training dataset the prepare the stage for the final, decisive transfer learning layers. 
Note that entering Section \ref{sec:exp-data}, we expand the dataset further by including more images from ImageNet.

\begin{figure}
\includegraphics[width=\columnwidth,clip]{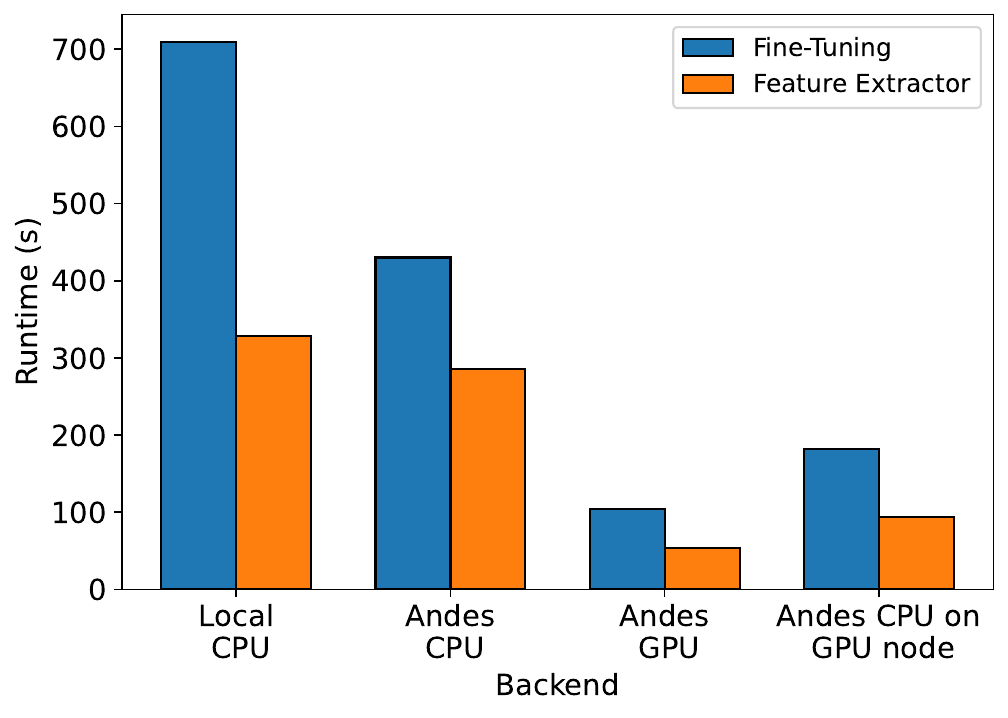}
\captionof{figure}{Changes in runtime due to the backend used and training method (Section II).}
\label{fig:timexback}
\end{figure}

We started with running the program on a local laptop to compare the final accuracy and runtimes for the two TL learning methods, "fine-tuning" the Convolution Network, and fixed feature testing.
Fine-tuning removes the final layer and adds in the pre-trained model from ImageNet, whereas the fixed feature extractor freezes the previous layers while finding the weights from the larger dataset\cite{fixed-feature}.
This method takes less time with a slight benefit to the final accuracy of the network (see Figure \ref{fig:timexback}).

Notice that the tests in this section were run on four different backends: an Andes CPU compute node, an Andes GPU node, an Andes CPU on a GPU node, and the local laptop used for the experiments, an Apple Macbook Pro (2.3 GHz Quad-Core Intel Core i7).
Andes\cite{Andes-compute-nodes} is a Linux Cluster located at OLCF. It hosts 704 batch (CPU) compute nodes, as well as 9 GPU compute nodes.
This program predictably ran much faster on Andes, particularly on the GPU node, regardless of the backend. 
However, due to the limited amount of GPU nodes on Andes, this approach is less feasible for larger training sets that require distributed training methods.

Once affirming that these tests could be run locally and remotely on Andes, we moved on to the hybrid program. 

\section{Running HPC-QC Hybrid Programs}

\subsection{Pathfinding} \label{sec:pathfinding}

In the second part of our work, we transitioned to a hybrid PennyLane\cite{pennylane} (v0.32.0) and PyTorch (v1.13.0) program using the same ants and bees subset from ImageNet.
PennyLane is a Python library, written and maintained by Canadian quantum photonics company Xanadu\footnote{https://www.xanadu.ai/}, which has built a reputation for reliably working with several quantum cloud backends, including those from IBMQ.
It also features a large repository of QML documentation, which we utilized for this project.

\begin{figure}
\includegraphics[width=\columnwidth,clip]{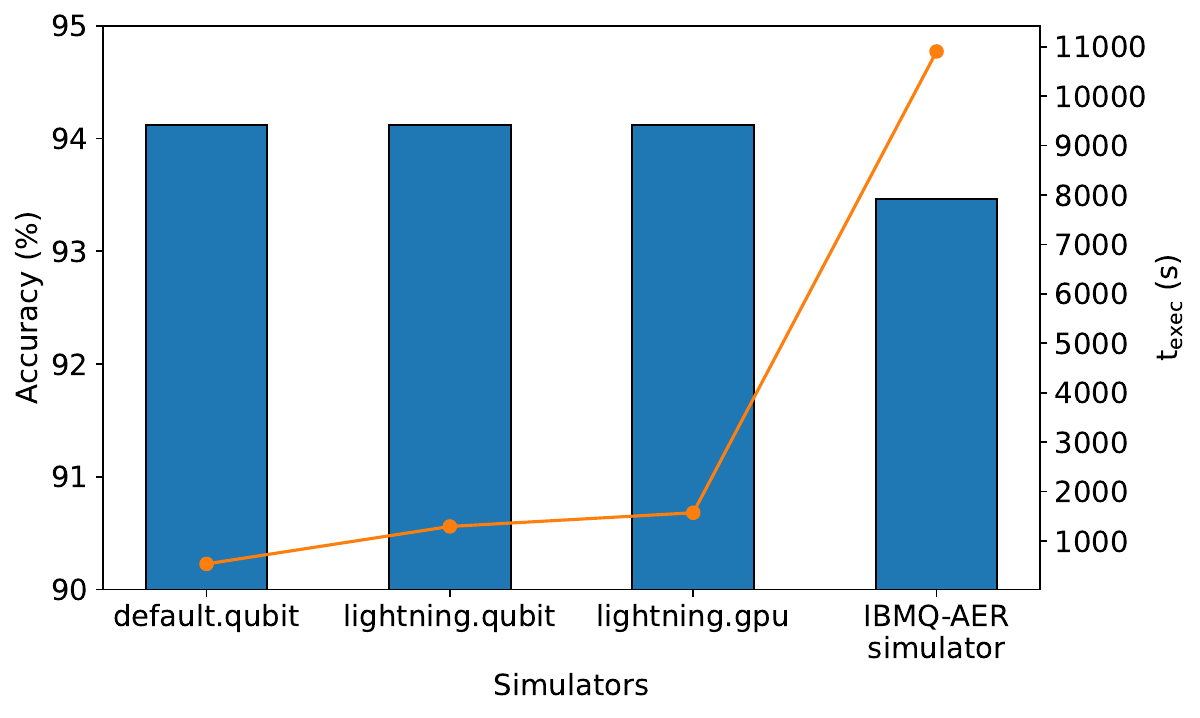}
\captionof{figure}{Accuracy (blue, left axis) and runtime (orange, right axis) versus simulator used. (4 qubits, 30 epochs on Andes)}
\label{fig:backends}
\end{figure}

\begin{figure}
\includegraphics[width=\columnwidth,clip]{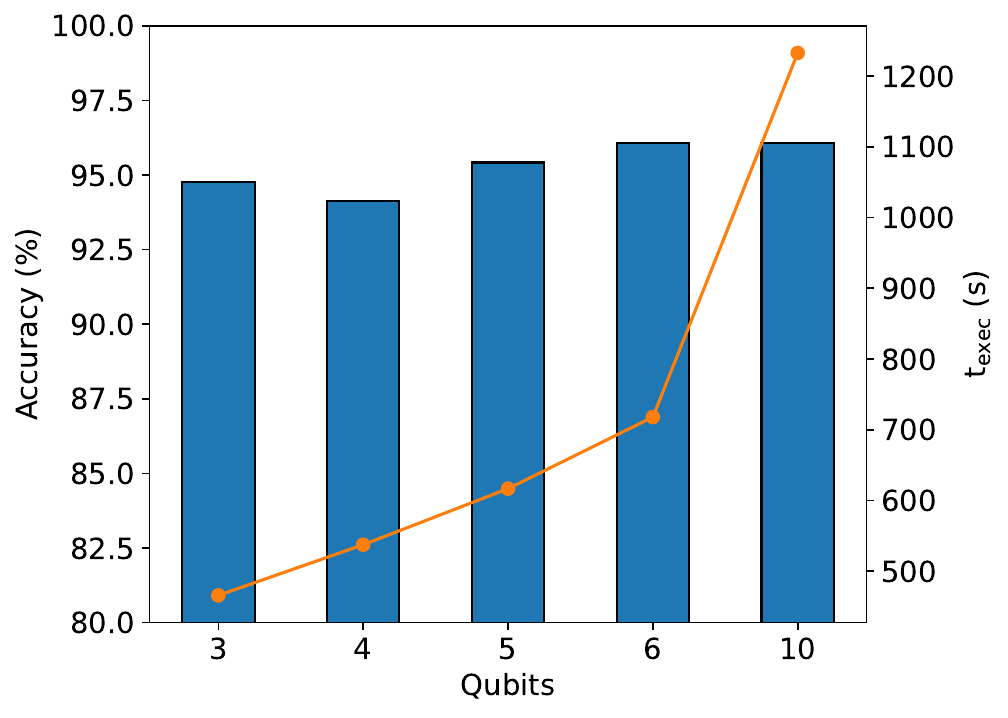}
\captionof{figure}{Accuracy (blue, left axis) and CPU runtime (orange, right axis) versus qubits. (30 epochs on Andes CPU)}
\label{fig:timeaccqubit}
\end{figure}

The hybrid program follows the same basic principles as the original PyTorch code but instead allows the quantum qubits to do most of the work.
After being initialized to quantum states depending on the original training input, they are put through 3D rotations and entanglements.
These allow the qubits to fundamentally follow the same procedures as their classical counterpart. 

Using the code from PennyLane\cite{pennylane-docs}, we set out to run the program while modifying various parameters to determine the most optimal way to balance the hybrid program.
Across all the classical compute systems we utilized, the three main code parameters explored were: 1) the quantum backend, 2) the epoch count, and 3) the qubit count.

From a quantum backend perspective, we first tested several of the most common quantum computing simulators to which PennyLane has access and compared the accuracy and runtime of the quantum circuit, as exhibited in Figure \ref{fig:backends}.
The three PennyLane simulators we tested, default.qubit, lightning.qubit, and lightning.gpu (run on an Andes GPU node) each had identical accuracy at approximately 94\%, with default.qubit having the shortest runtime by a significant margin.
However, switching to other resources caused the runtime and accuracy to plummet in efficiency.
Although the lightning.gpu did not perform well on Andes, we suspect it would have performed better when scaling across multiple GPUs and when used on a more GPU-centric machine (such as Frontier); however, Frontier is unable to run the lightning.gpu device due to lack of AMD support.
Both the lightning and lightning.gpu devices are also known to perform better with larger qubit counts than the 4 qubit tests we performed \cite{O’Riordan_2022}.

\begin{figure}
\includegraphics[width=\columnwidth,clip]{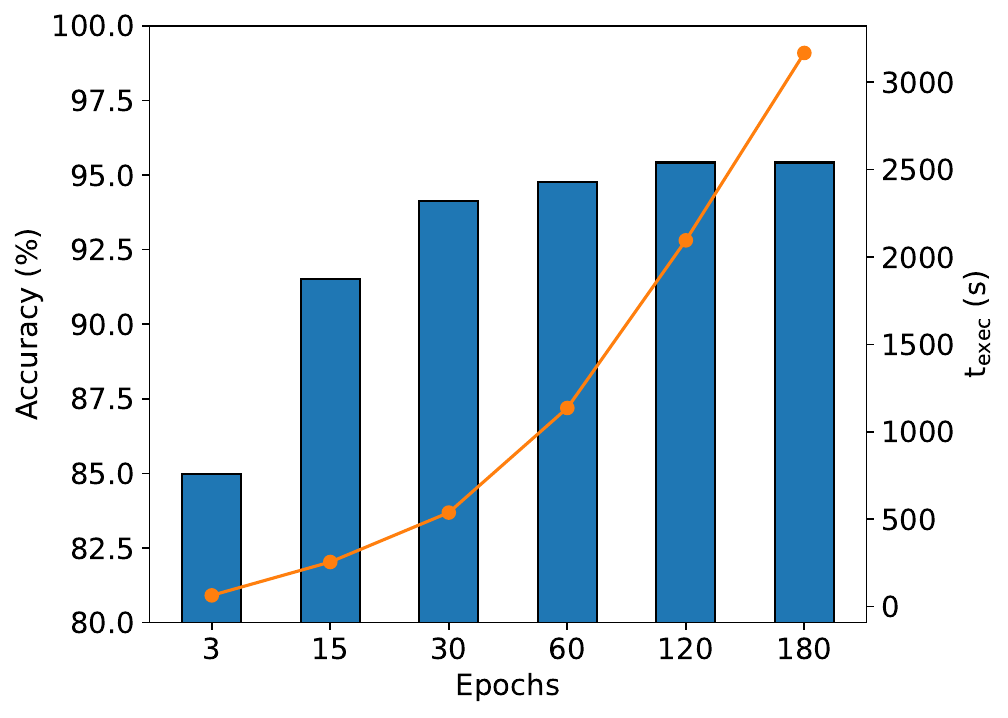}
\captionof{figure}{Accuracy (blue, left axis) and CPU runtime (orange, right axis) versus total epochs. (4 qubits on Andes CPU)}
\label{fig:timeaccepoch}
\end{figure}

Exploring outside of PennyLane simulators, we tested IBM's AER simulator \cite{qiskit2024}.
Accessed locally, it still took nearly ten times longer than the default PennyLane simulator, and it operated at a less stable and lower accuracy of approximately 93\%. Switching to devices accessed over the Cloud, we attempted jobs on IBMQ's QASM simulator \cite{qiskit2024}. 
Unfortunately, we determined that this was not possible, as the number of jobs the program needed to submit to the simulator was far too much to work on one node of Andes feasibly.
Each time we exceeded the apparent IBMQ limits, as jobs started to fail approximately 5 hours in (18,000-19,500 seconds), or slightly less than one epoch through. 
The training algorithm used performs thousands of tensor-based (or array-based) calculations, and by relying on the quantum computer for each of them, the runtime is significantly increased. 
IBM also uses a queue for their QASM simulator, like their quantum computing devices, which though short (1-5 seconds), prevents the ML algorithm from completing in a reasonable amount of time.
We additionally tried to run on other IBMQ backends before realizing the unfeasibility of this as well. 
The average remote IBMQ queue time for a user at OLCF is close to 10 hours for one job, and due to the fair-share queue policy\cite{fair-share}, it can increase for everyone in the project if many jobs are submitted.
Thus, we deemed that this QML training program could not run in a computationally reasonable amount of time on a remote quantum computer (at least in our code's \textit{current} state).

The remaining code parameters that we independently adjusted were the epochs and qubits.
Epochs define how many times the training iterates through the algorithm, increasing accuracy at the expense of runtime\cite{epoch-define}, while qubits are the basic representation of encoded quantum data.
When increasing the number of qubits, the same limitations are present as increasing epochs.
It is worth noting that some of the most powerful quantum processors today range from 30-430 qubits\footnote{https://quantum-computing.ibm.com/services/resources?tab=systems}, but often feature long queue times waiting for jobs to run, so we wanted to test and determine how detrimental this would be for a basic QML program.

While keeping the qubits at the default value, 4, we increased epochs until we saw a significant increase in runtime without a change in accuracy.
This occurred at approximately 120 Epochs, as per Figure \ref{fig:timeaccepoch}.
It is worth noting that the recommended 30 Epochs operate 4 times faster with a 4\% decrease in accuracy, confirming that this could also be a sufficient benchmark for the variation in simulated qubit count.

The qubits offered less expected results.
Decreasing the qubit count to 3 (from the default 4) decreased runtime and increased accuracy.
Increasing qubit count steadily increased the accuracy to a plateau at approximately 96\%, but after 6 qubits, there was a greater gradient concerning runtime (Figure \ref{fig:timeaccqubit}).
Simulating with much more than 10 qubits (15-20) did not increase accuracy at all, and runtime only continued to exponentially increase until stopping at 20, where errors prevented the program from running fully.

Note that, until this point, we were purely invoking the Python program via \texttt{ python script.py } and not explicitly mapping to a specific number of cores or threads on any system.
In the next section we transition to distributed training, thus enabling us to take more advantage of a given system (especially the HPC systems).

\subsection{Parallelization} \label{sec:parallel}

\begin{figure}
\includegraphics[width=\columnwidth,clip]{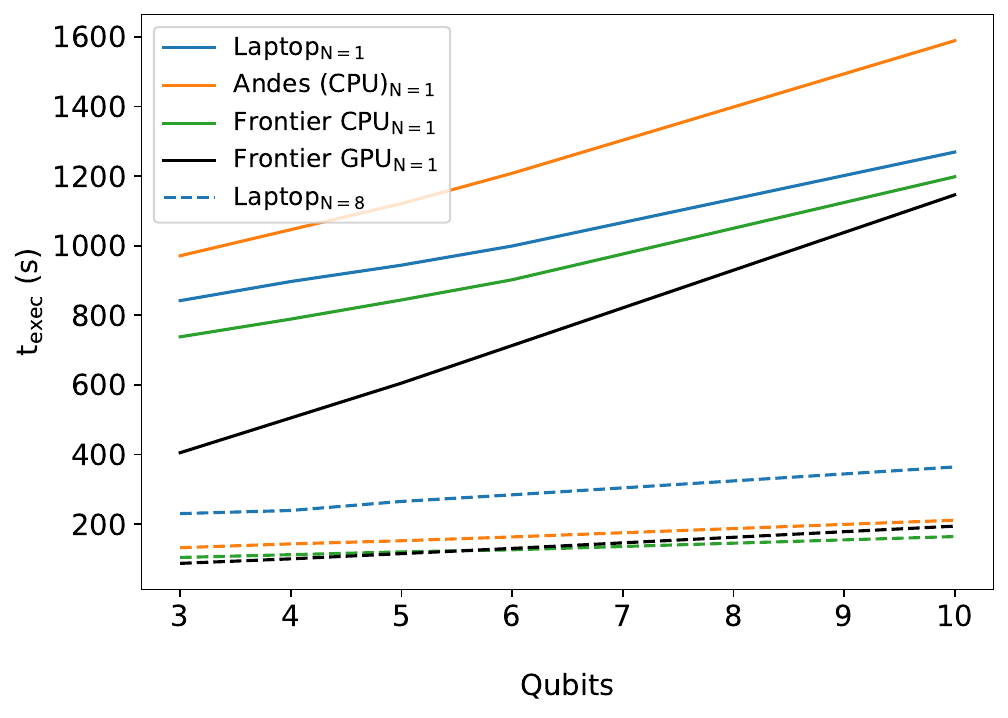}
\captionof{figure}{Qubit runtime by system. Execution time for $\rm N=1$ (solid) and $\rm N=8$ (dashed) threads across all systems.}
\label{fig:cpugpulaptop2}
\end{figure}

The third part of our work focused on distributing the QML training across multiple threads -- either across multiple CPU cores or GPU devices.
For this suite of tests, we added Frontier\cite{Frontier-compute-nodes} -- the world's first exascale computer -- into our list of test systems.
PyTorch is commonly utilized on NVIDIA GPUs, meaning that a different build method was needed to run the code on Frontier's AMD\footnote{https://docs.olcf.ornl.gov/systems/frontier\_user\_guide.html\#amd-gpus} GPUs.
Although pre-compiled binaries are available with AMD support, we noticed a slight performance increase on Frontier when building from source instead.
Implementing the QML code required combining what we already had used with a Distributed Data-Parallel (DDP) training method.
Using DDP and a distributed sampler, we were able to implement a workflow that allowed the code to train across multiple CPUs, GPUs, and compute nodes.
Note that due to the small ants and bees dataset, multiple compute nodes were not needed in this scenario, but are needed as the dataset is increased (see Section \ref{sec:exp-data}).
While we explored both CPU and GPU distribution on Frontier, CPU training was the focus on Andes due to the limited nature of the GPU compute nodes.
We also tested a similar workflow on the laptop with its multiple compute cores. 
These tests were conducted using the default.qubit simulator from Pennylane to yield results that are more comparable to the previous ones.

For all systems used (see Figure \ref{fig:cpugpulaptop2}), we tested the performance of the code when using $\rm N=$ 1 thread (effectively non-distributed training) compared to $\rm N=$ 8 threads.
The 8 threads are a convenient choice due to the laptop's maximum thread count (4 cores, two threads per core) and the maximum number of GPUs per node on Frontier (4 GPUs, with 2 GCDs per GPU).
For our purposes, all GCDs on Frontier will be referred to as "GPUs".
On the HPC systems, distribution was achieved through the \texttt{srun} parallel job launcher, while PyTorch's native \texttt{torchrun} was used instead for the laptop \cite{slurm}.
For tighter control over thread mapping on the HPC systems, MPI was utilized via the \texttt{mpi4py} package \cite{mpi4py}.

\begin{figure}
\includegraphics[width=\columnwidth,clip]{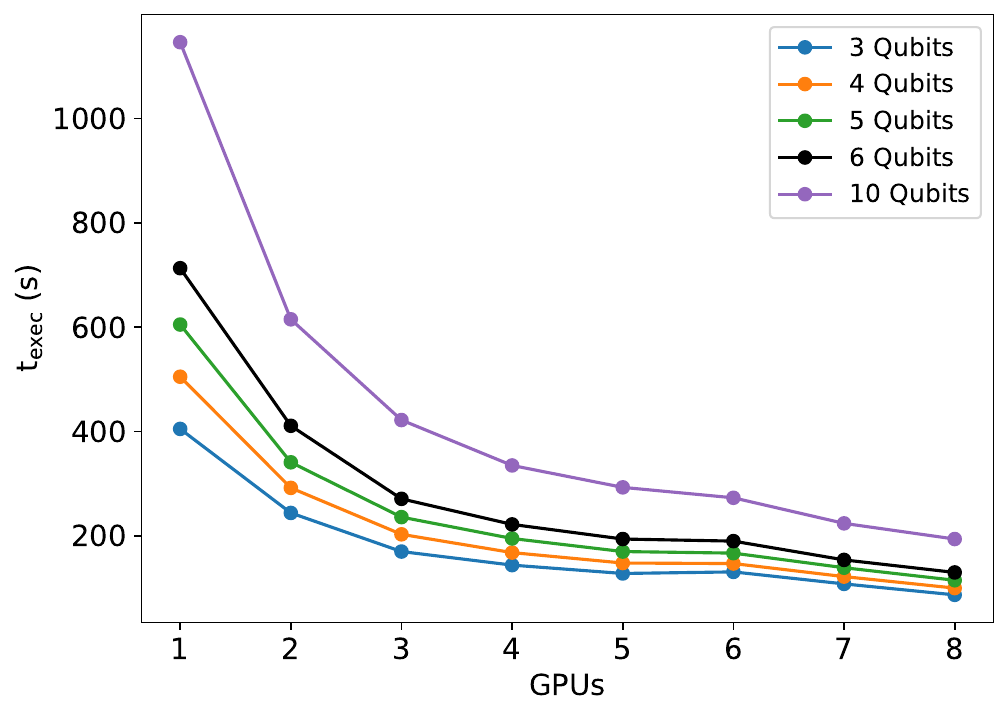}
\captionof{figure}{Execution time by qubit count and number of GPUs on Frontier. (30 epochs, smaller dataset)}
\label{fig:gpuxqubit}
\end{figure}

As expected, the laptop performs well when only considering one compute core (even outperforming Andes), but falls short when starting to distribute across multiple cores.
Frontier's GPU performs the best in the $\rm N=1$ regime, but has a trend that will eventually be outperformed by the Frontier CPU at larger qubit counts.
When considering $\rm N=8$, Frontier's CPU starts to perform quicker than the GPU at 6 qubits.
The loss of performance at larger qubit counts is a bit concerning, as the slope of the Frontier GPU performance is drastically different when compared to other methods.
This may be attributed to GPU communication overheads in the quantum circuit every training iteration (which worsens as the number of qubits is increased), or an underlying communication issue between CPU and GPU torch devices in the circuit due to using the default.qubit simulator.
This may be mitigated by using the tailored lightning.gpu simulator; however, it is currently unable to be installed on Frontier due to hardware compatibility issues.
The new lightning.kokkos simulator, which should be compatible, was also not an option at the time of testing.
Due to the code hanging at certain GPU counts and not hanging when running purely on the CPU with an equivalent number of threads, an underlying GPU communication issue with the code seems highly likely.
One other aspect is that our code was never designed to be run at larger qubit counts and is thus inefficient in that region.
Even with Frontier's GPU trend being eventually surpassed by the Andes CPUs at larger qubit counts, all HPC systems maintain better performance (including their slopes) than the laptop approach at $\rm N=8$.
The laptop results emphasize that although one can run this program without HPC resources, laptop performance quickly falls behind as more and more threads are needed for distributed training (further emphasized with a larger dataset in Section \ref{sec:exp-data}).

\begin{figure}
\includegraphics[width=\columnwidth,clip]{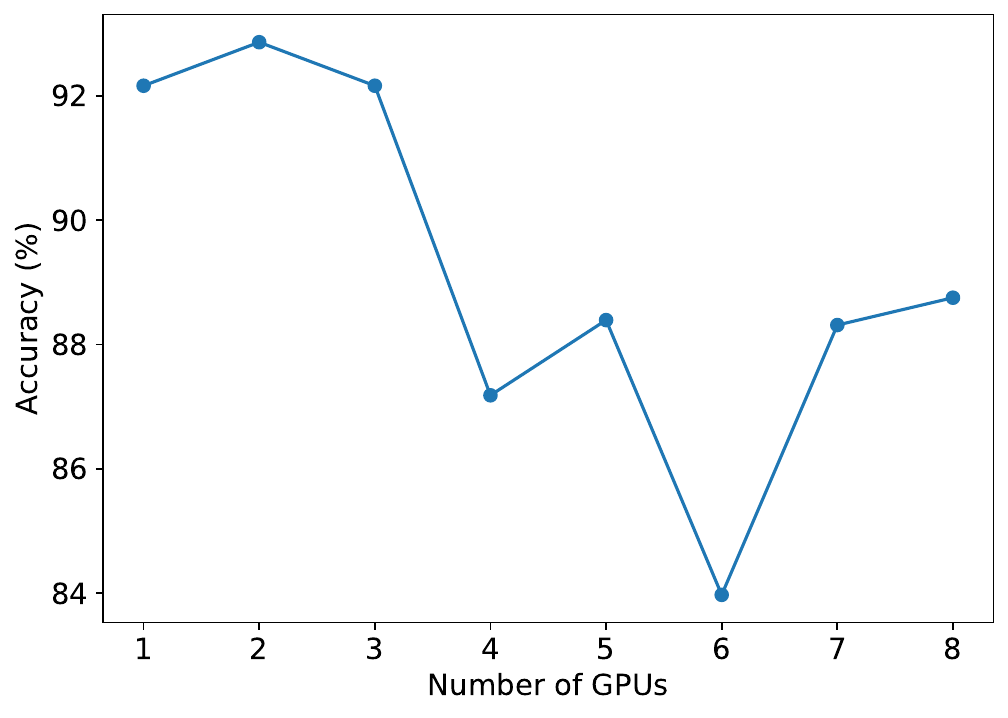}
\captionof{figure}{Accuracy of the smaller dataset versus number of GPUs on Frontier. (4 qubits, 30 epochs, smaller dataset)}
\label{fig:gpuxacc}
\end{figure}

Note that the discrepancy between the Andes times listed in Figure \ref{fig:cpugpulaptop2} and Figure \ref{fig:timeaccqubit} are due to now explicitly invoking 1 or 8 threads via \texttt{srun} as opposed to purely doing \texttt{ python script.py }.
The former leads to more consistent and precise optimization, while the latter approach had inconsistent optimization across compute cores and is limited to 1 compute node.

Focusing more on Frontier's performance, we benchmarked a single compute node across our qubit range.
As exhibited in Fig. \ref{fig:gpuxqubit}, the most optimal execution of the circuit, in terms of runtime, is on 8 GPUs with 3 qubits.
When utilizing the full power of a compute node on Frontier with all 8 GPUs, the runtime plateaus across all qubit counts.
This is explained by the number of training iterations being small due to the relatively small dataset, so adding more GPUs to distribute the training provides less benefit.
Note that at larger thread counts, the distributed training masks the effect that larger qubit counts have on runtime, while the quantum portion of the computations has more of an impact at lower thread counts.

One thing to consider is that other hyperparameters, such as batch size and layers, also present issues. 
In Section \ref{sec:pathfinding}, we used the default circuit depth of 6 and batch size of 4, and we used these same benchmarks for consistent comparative testing in Section \ref{sec:parallel}; however, in reality, these are both very small values.
CNNs often have a base of 15 layers but at this point the Pennylane default.qubit simulator would have significant trouble computing all of the rotations necessary on the input qubits. 
Remote backends such as those used by IBMQ and Quantinuum may not be able to run those circuits with success either due to their depth.
The best quantum computers on the market have high accuracy at a circuit depth range of only 10-20\cite{Q-Volume}.
This means that any large QML datasets are best trained on simulators until stronger, more coherent quantum computing resources are available, or when better algorithms are implemented\cite{depth}. 

Likewise, batches determine how often the model updates, and larger dataset models often use anywhere from 32 to 'the size of the training set' batches\cite{batch}.
If we tested with a larger dataset, these bounds would be more appropriate.
This is most significantly reflected in the accuracy of the training falling while more GPUs are added to the workflow, as exhibited in Fig. \ref{fig:gpuxacc}.
While accuracy is maintained in the early stages from 1-3 GPUs, it plummets as more are added and the sampler splits the dataset into smaller and smaller parts.
Due to keeping the batch size parameter constant, the "effective batch size" (the product between the number of GPUs and batch size per GPU) is directly proportional to the increase of GPUs.
The effective batch size increasing, combined with not increasing the learning rate parameter accordingly, results in inconsistent, low image recognition accuracy.

\subsection{Expanding the dataset} \label{sec:exp-data}

In an effort to analyze performance across multiple compute nodes, we expanded the dataset to include more images from ImageNet.
The overall dataset was expanded to 4145 training images (up from 245), and also added an additional class (ladybugs).
Although ladybugs were added to the ants and bees set, the small amount of classes still keeps the model in the realm of a "feature extractor".
For these tests, we keep the number of qubits (4) and epochs (30) constant as we focus on analyzing performance of the default parameters on a larger dataset.
However, in an effort to combat the issue mentioned at the end of Section \ref{sec:parallel}, we scaled the learning rate parameter appropriately when distributing across more GPUs or CPUs\cite{2015deepresidual,1hourimagenet}.
Varying the learning rate in this manner led to a relatively consistent validation accuracy of $\sim$92\% as we scaled up the number of devices being used.


From a computational setup point of view, each MPI task was allocated the maximum number of physical cores in each L3 cache region on a given node of a system.
This meant spanning each MPI task across two compute cores on Andes, while spanning across seven compute cores on Frontier.
Mapping tasks to cores in this manner leads to a maximum of 16 tasks per node on Andes, and 8 tasks per node on Frontier.
Similar to Section \ref{sec:parallel}, CPU training was the focus on Andes, while both CPU and GPU training was explored on Frontier.

\begin{figure}
\includegraphics[width=\columnwidth,clip]{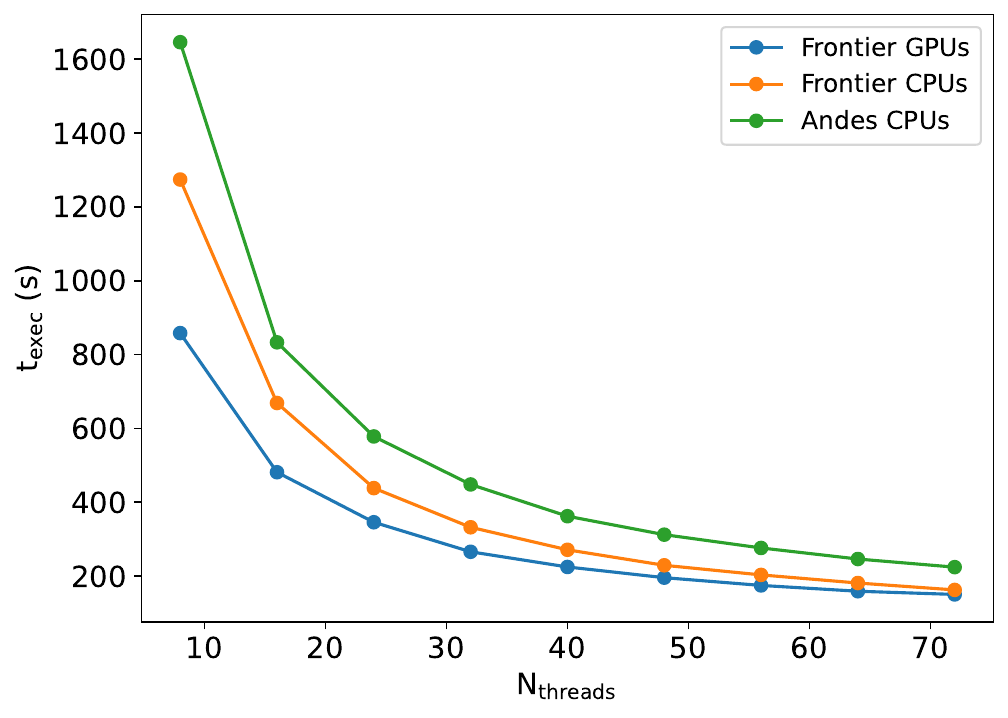}
\captionof{figure}{Execution time versus number of threads on Andes and Frontier. (4 qubits, 30 epochs, larger dataset)}
\label{fig:frontier_super}
\end{figure}

The performance of our hybrid program across multiple compute nodes can be seen in Figure \ref{fig:frontier_super}.
As expected, the Frontier GPUs perform quite well overall, performing $\sim$92\% and $\sim$48\% faster than the Andes and Frontier CPUs, respectively, at $\rm N=8$. 
Not pictured in Figure \ref{fig:frontier_super} is the laptop's performance on the larger dataset using 8 threads, which had an execution time of 2801~s.
This leads to an astounding speed-up of $\sim$226\% that the Frontier GPUs have over the laptop at the same thread count.
The Frontier CPUs and GPUs start to converge around 72 threads (9 nodes), where the GPUs are only $\sim$8\% faster.
Note that although not constant, the percent difference between the Andes CPUs and Frontier CPUs is consistently around the 30\% region.
Like in Section \ref{sec:parallel}, this suggests inefficient GPU communication in our code that degrades the Frontier GPU tests as thread count is increased; although, this time it was not the high qubit count but rather the shear number of GPUs being used.
With the absence of a GPU-based simulator, this problem will only worsen as the dataset is expanded further, or when more nodes ($\geq$10) are used with this dataset. For further details and access to the code, please refer to the GitHub link provided below\footnote{https://github.com/michael-sandoval/hybrid\_quantum\_hpc}.

Although Andes clearly performs the slowest when compared to both Frontier methods, it is able to use twice as many tasks per node due to the way we mapped MPI tasks to cores (1 MPI task per L3 memory region), thus providing a potential benefit to cost when comparing to the Frontier CPU method.
However, more testing must be done with the Frontier CPU approach, as Frontier has twice as many physical cores per node than Andes, but utilizing more tasks per node on Frontier would require packing tasks more tightly in a given L3 region, which could affect performance.

\nocite{Yang_2019} 
\nocite{gpu-ml} 

\section{Discussion and Conclusions}

Although ML has proven its value in the realm of HPC, this work sought out to explore the potential value QML has in an HPC environment from a quantum simulator standpoint.
Taking a "ground up" approach, we started with a tutorial code and gradually modified it and the dataset to fit in an HPC environment.
We highlighted the HPC benefits that QML intrinsically inherits from classical ML (data distribution speed-up), while also demonstrating potential quantum benefits (model accuracy at higher qubit counts) and simulator pitfalls.
One result we emphasize is the communication obstacles when using either cloud-based or local simulators.
More specifically, the latency to remote, cloud-based simulators is concerning when simulators are used every training iteration.
Additionally, inefficient GPU communication with local, CPU-based simulators pose a challenge \textit{within} and \textit{across} compute nodes at increased qubit and thread counts, respectively.
The communication challenges emphasize how a simulator should be implemented, including its framework, and ideally how it should be designed with HPC in mind\cite{xacc}.

This research furthers the active interest in hybrid QC and HPC resources located in the same location at computing centers worldwide.
While QC and HPC can operate independently, remote access, particularly for job-partitioned QML algorithms, is best suited for well-developed workflows in computing centers.
Some suggestions are noted as follows:

\begin{enumerate}[label=(\Alph*)]
    \item Run jobs on devices featuring both classical and quantum computational abilities: QPUs, CPUs, and/or GPUs. 
    \item Port jobs to centers hosting both classical and quantum computers at the same location\cite{meetiqm}.
    \item Reserve time on HPC and quantum devices (for the same period) ahead so that the full power of the QC can be utilized during HPC runtime without queue-induced latency. 
    \item Design QML jobs that can be trained in parallel with multiple GPUs/nodes with a GPU-capable simulator (in addition to any of the above).
\end{enumerate}

Further research can also be done with larger datasets to determine the feasibility of hybrid programs.
Many large databases such as ImageNet require strong CPU or GPU infrastructure, but other datasets, particularly those used in the private sector for demographic mapping\cite{google}, self-driving vehicles\cite{self-driving}, and forensic recognition\cite{digital-forensics}, may require large server architectures or many HPC machine nodes to train \cite{IT-KB-NDSU}.
QML would be especially optimal for these use cases, offering not only speedup but a practical application for early quantum computers in the Noisy Intermediate Scale Quantum era. 

Additionally, this research acts as an example for the benefit an HPC environment can provide to quantum computing -- even when only considering quantum simulators (as opposed to a \textit{real} quantum backend).
Local non-HPC machines can run simulators adequately in some scenarios, but quickly fall behind when multiple threads are needed to perform computations (i.e., with complex training methods or when using large datasets).
In Section \ref{sec:parallel}, the laptop is consistently outperformed in most scenarios, even when utilizing a small dataset, as demonstrated.
This trend is reinforced in Section \ref{sec:exp-data}, where the impracticality of using the laptop becomes apparent as we increase the dataset size.
Although there are much more powerful non-HPC machines available when compared to the laptop we used in this research, the takeaway remains that they will eventually hit feasible training limits much earlier than a compute cluster will (even when only using a simulator).
We acknowledge that this research was based on a tutorial-like problem, but hope that this proof of concept will motivate more intensive scaling studies in the future.


\begin{acknowledgments}

This research was supported in part by an appointment to the Oak Ridge National Laboratory Science Undergraduate Laboratory Internship Program, sponsored by the U.S. Department of Energy and administered by the Oak Ridge Institute for Science and Education.
STB thanks Art Stewart for paper reviewing services. 
We acknowledge the use of IBM Quantum services for this work.
The views expressed are those of the authors, and do not reflect the official policy or position of IBM or the IBM Quantum team.
This research used resources of the Oak Ridge Leadership Computing Facility at the Oak Ridge National Laboratory, which is supported by the Office of Science of the U.S. Department of Energy under Contract No. DE-AC05-00OR22725.
We thank Oak Ridge National Laboratory Leadership Computing Facility for allowing us access to Quantum Processors at IBMQ, as well as the Andes Linux Cluster and Frontier Exascale computers. 

\end{acknowledgments}

\bibliographystyle{ieeetr}
\section*{References}
\bibliography{references.bib}
 

\end{document}